\documentclass[rog]{agutex}

\usepackage{graphicx}
\authorrunninghead{D. B. JESS \& G. VERTH}

\titlerunninghead{MHD Waves in the Solar Photosphere}

\begin{document}

%
%

\title{Ultra-high-resolution Observations of MHD Waves in Photospheric Magnetic Structures}
%
%

%
%



\authors{David B. Jess\altaffilmark{1} and Gary Verth\altaffilmark{2}}

\altaffiltext{1}{Astrophysics Research Centre,
School of Mathematics and Physics,
Queen's University Belfast, Belfast, Northern Ireland, 
BT7 1NN, UK.}

\altaffiltext{2}{
Solar Physics and Space Plasma Research Centre (SP$^{2}$RC),
The University of Sheffield,
Hicks Building, Hounsfield Road, 
Sheffield, S3 7RH, UK.}





%
%


\begin{abstract}
Here we review the recent progress made in the detection, 
examination, characterisation and interpretation of 
oscillations manifesting in small-scale magnetic elements 
in the solar photosphere. 
For whitelight observations, the solar photosphere is defined 
as the layer that emits most of the Sun's electromagnetic radiation, 
and therefore creates the limiting depth of optical imaging sequences.
This region of the Sun's atmosphere is especially 
dynamic, constantly evolving, and 
importantly, permeated with an abundance of 
large- and small-scale concentrations of magnetic 
fields. Such magnetic features can span 
diameters of hundreds to many tens of thousands of km, 
and are thus commonly referred to as the `building blocks' 
of the magnetic solar atmosphere. However, it is the 
smallest, and by far the most numerous magnetic elements 
that have risen to the forefront of solar physics 
research in recent years. Structures, which include magnetic 
bright points, are often at the diffraction limit of 
even the largest of solar telescopes. Importantly, it is the 
improvements in facilities, instrumentation, imaging techniques 
and processing algorithms during recent years that have allowed 
researchers to examine the motions, 
dynamics and evolution of such features on the smallest 
spatial and temporal scales to date. It is clear that while these structures 
may demonstrate significant magnetic field strengths, their 
small sizes make them prone to the buffeting supplied by the 
ubiquitous surrounding convective plasma motions. 
Here, it is believed that magnetohydrodynamic 
waves can be induced, which propagate along the field 
lines, carrying energy upwards to the outermost 
extremities of the solar corona. Such wave 
phenomena can exist in a variety of guises, including 
fast and slow magneto-acoustic modes, in addition to 
Alfv{\'{e}}n waves. Coupled with rapid advancements in
magnetohydrodynamic wave theory, we are now in
an ideal position to thoroughly investigate how wave motion 
is generated in the solar photosphere, which oscillatory 
modes are most prevalent, and the role that these waves 
play in supplying energy to various layers of the solar 
atmosphere. 
\end{abstract}

%
%

%

\begin{article}

%
%

\section{Introduction}
Ever since oscillatory motion was first discovered in the Sun's 
tenuous atmosphere \citep{Lei60, Lei62, Noy63}, it has been 
a goal amongst physicists to detect, identify, characterise and 
understand the diverse variety of wave modes manifesting 
in the solar atmosphere. Initial observations in the optical portion 
of the electromagnetic spectrum allowed oscillations 
present in the photospheric and chromospheric layers to be 
studied. However, during these early stages technology was 
not as advanced as it is today, and as a result, key modern 
techniques such as adaptive optics \citep[AO; e.g.,][]{Rim11}, 
multi-object multi-frame blind deconvolution \citep[MOMFBD;][]{Van05} 
and speckle reconstruction \citep{Wog08} were unavailable 
to help combat the fine-scale image degradation caused by the Earth's 
atmosphere. Therefore, initial research was dedicated to probing 
large-scale solar structures, including sunspots and super-granules. 
However, even at these large spatial scales, a wealth of 
oscillatory phenomena were found to be omnipresent 
\citep[e.g.,][]{Deu69, Ulr70, Deu71}. Such waves 
demonstrated periodic intensity and velocity fluctuations, and 
thus were placed under the same umbrella as acoustic modes, 
which are ultimately defined by their intrinsic signatures of 
compressions and rarefactions.
Difficulties were encountered in follow-up work when the 
measured phase velocities of the waves were found to 
be too large to be explained by over-simplistic acoustic 
modelling. Instead, \citet{Ost61} and \citet{Mei76} 
hypothesised that the magnetic fields in which the waves 
were embedded within must also be taken into consideration. 

Over the following decades the examination of magneto-acoustic 
waves rose to the forefront of observational solar physics, 
with \citet{Ulm76} aptly asking whether they may provide a 
significant channel for energy leaking into the outer layers 
of the Sun's atmosphere. Furthermore, the mere presence of a magnetic 
field also introduces a number of additional viable wave modes that 
may simultaneously exist within the atmosphere. Through 
magneto-hydrodynamic (MHD) modelling it was found that although 
these additional waves have similarities with purely acoustic modes, they 
are often highly anisotropic. This is because the addition of the 
magnetic field introduces a dependency on both ({\it{i}}) the alignment of the 
wavevector, $k$, with the direction of the background magnetic field, $B_{0}$, 
and ({\it{ii}}) the ratio of the kinetic pressure, $p_{0}$, to the magnetic pressure, 
$B_{0}^{2}/2{\mu}_{0}$ (or $B_{0}^{2}/8{\pi}$ in cgs units),
in the environment which supports the wave.
This ratio is commonly referred to as the plasma $\beta$, defined as 
$\beta = 2{\mu}_{0}p_{0}/B_{0}^{2}$, where ${\mu}_{0}$ is the magnetic 
permeability of free space. This quantity can be re-written in terms of the 
local Hydrogen number density, $n_{H}$, the plasma temperature, $T$, 
and the Boltzmann constant, $k_{B}$, as $\beta = 8{\pi}n_{H}Tk_{B}/B_{0}^{2}$, 
providing a representation of the plasma $\beta$ in cgs units.
Thus, in the lower layers of the solar atmosphere where the temperature is 
relatively low ($T\sim6000$~K) and the magnetic field strength is still 
intensely concentrated ($B_{0}>1000$~G), it is clear to see how a 
majority of the magnetic structures supporting magneto-hydrodynamic wave 
phenomena are often defined by $\beta{\ll}1$ (i.e., dominated by magnetic 
pressure). This has important consequences for the wave modes 
that are expected to exist within this plasma regime, and allows for the 
manifestation of `fast' and `slow' magneto-acoustic waves, in addition to 
Alfv{\'{e}}n waves \citep{Edw83}. Structures in the lower solar atmosphere 
which demonstrate oscillatory behaviour are often observed to be 
long, elongated features (e.g., mottles, spicules, fibrils, etc.), 
and as such are typically modelled by employing cylindrical 
geometry. Here, an overdense magnetic `flux tube' will create an efficient 
waveguide that will support an even richer variety of MHD waves depending 
on the azimuthal wavenumber, $m$, including 
sausage ($m=0$), kink ($m=1$) and fluting modes ($m>1$), which are yet 
further classified via their trapped/leaky, fast/slow and body/surface 
characteristics. Thus, the quest to detect all of the various MHD 
wave modes has intensified, but so too has the drive to determine the precise 
role they play in the transportation and dissipation of energy through 
the Sun's atmosphere. 

The ongoing research into waves and oscillations in the solar 
atmosphere does not take place for solely esoteric purposes.
Rather, it has the ability to allow us to delve deep below the visible 
solar surface through helioseismology techniques \citep[e.g.,][]{Duv93, Sch98, Lop14}, 
and to understand the coupling between (quasi-) periodic flows
and plasma motions that are 
abundantly apparent over a wide range of spatial scales 
(e.g., spicules, mottles, fibrils, plumes, prominences, etc.). 
Quantifying the possible contribution of observed MHD 
waves to plasma heating is also of immense 
interest in solar physics. 
The paradoxical nature of how the Sun's outer atmosphere 
is heated 
to (and maintained at) multi-million degree temperatures 
is a problem that has been plaguing scientists for over 
half a century. Even the solar chromosphere, a thin 
atmospheric layer that is only heated to a 
few thousand degrees above 
the underlying photosphere, requires 
extraordinary plasma heating 
processes to balance the radiative losses experienced 
in this relatively high-density environment. It has long 
been believed that MHD waves, generated in the photosphere 
and channelled upwards along magnetic field lines, 
may be able to contribute directly to plasma heating providing a 
suitable (and efficient) conversion mechanism exists.

\begin{figure*}
\begin{center}
\includegraphics[angle=0,width=16cm]{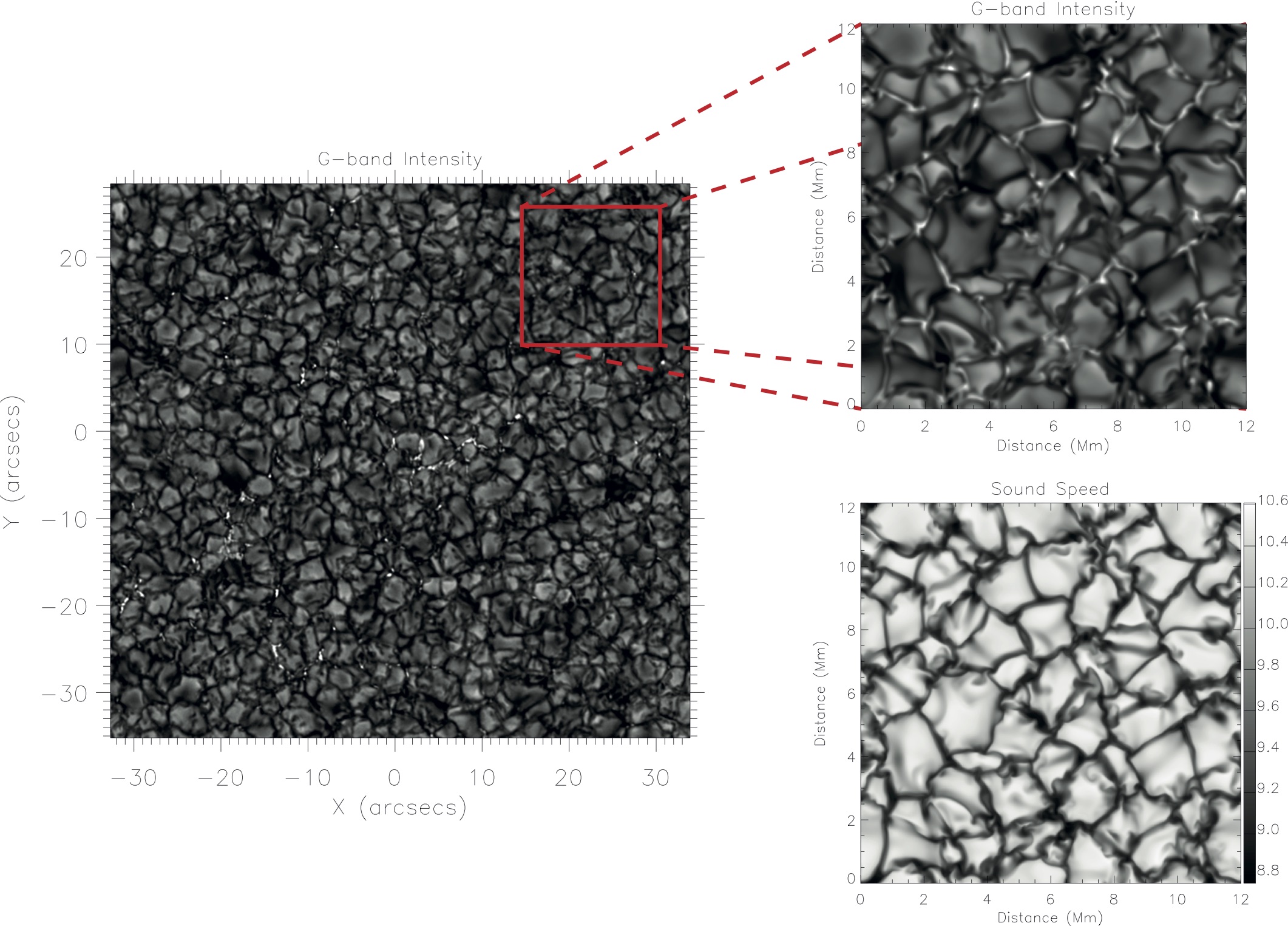}
\end{center}
\caption{A G-band image (left), acquired by the ROSA multi-camera system 
at the Dunn Solar Telescope, revealing a large number of 
MBPs. These features are visible 
as intensity enhancements within the intergranular lanes, and 
often demonstrate magnetic field strengths exceeding 1000~G. The axes 
are in heliocentric arcseconds, where $1''\sim0.725$~Mm. The red square 
highlights a $12\times12$~Mm$^{2}$ subregion of the field-of-view, with 
an equally-sized simulated G-band image displayed in the upper-right panel. 
The output of the MuRAM numerical code also reveals bright 
magnetic features manifesting within the intergranular lanes, suggesting a significant 
amount of agreement between current high-resolution observations and 
radiative magnetohydrodynamic modelling techniques. The 
lower-right panel displays the local sound speeds calculated from the 
local densities and pressures in the simulated polytropic atmosphere. It is clear that 
intergranular lanes, where magnetic bright points reside, often display sound speeds 
on the order of 10~km{\,}s$^{-1}$. Images adapted from \citet{Jes12b}.}
\label{Jess_MBP_figure}
\end{figure*}

To generate localised heating from MHD waves, 
smaller length scales must be created via physical processes 
such as resonant absorption, phase mixing or shock formation. 
Observing such fine-scale plasma dynamics directly presents a 
significant observational challenge due to the near- (or sub-) 
resolution scales involved.
\citet{DeP04} demonstrated how ubiquitous $p$-mode 
oscillations can be channelled along magnetic field lines, 
ultimately giving rise to dynamic phenomena in the chromosphere and 
transition region. Furthermore, \citet{Han06} revealed how 
shocks can form when slow magneto-acoustic waves, which are 
generated by convective flows and global $p$-mode oscillations, 
leak upwards from the solar surface along magnetic field lines 
and encounter the steep density gradients intrinsic to the chromosphere.
Of significant importance is the fact that there is now overwhelming 
evidence to suggest that waves have the ability to deform magnetic 
field lines and induce the necessary instabilities required to incite 
reconnective phenomena over a wide range of atmospheric heights 
\citep[e.g.,][]{Iso06, Iso07, Jes10, Li12, Jac13, She14}. Thus, 
examining the generation and behaviour of oscillatory phenomena 
in the solar atmosphere has the potential to shine light on a wide 
variety of physical phenomena, from magnetic reconnection on 
sub-arcsecond scales through to prominence eruptions spanning 
many hundreds of arcseconds.

In more modern times, the advent of high-sensitivity and low-noise 
camera systems has allowed short-exposure and high-cadence image 
sequences to be obtained from both ground- and space-based 
observatories. In particular, for ground-based solar telescopes 
such as the Dunn Solar Telescope (DST) and the Swedish 
Solar Telescope (SST), the successful application of adaptive 
optics and post-processing techniques has allowed researchers to focus 
their attention on the smallest magnetic elements observed in the 
Sun's lower atmosphere. Features that are close to the 
resolution limits of current solar telescopes, including magnetic 
bright points \citep[MBPs;][]{Dun73, Ste85, Sol93}, offer unique 
advantages over more large-scale structures (e.g., sunspots) 
in the study of MHD wave phenomena. 
For instance, these omnipresent elements, easily identifiable 
and ideal for feature tracking algorithms, are not as magnetically 
complex as macro-sized structures such as sunspots, 
and are therefore more readily compared with cylindrical 
geometry MHD approximations.
Furthermore, due to their small size 
\citep[often less than 0.3$''$ or 220{\,}km in diameter;][]{Cro10}, 
these structures are more prone to the buffeting imposed by the 
complex evolution of surrounding granules, thus increasing the 
likelihood of generating wave motion at the photospheric 
base of the magnetic field lines. In addition, the magnetic field 
strengths associated with MBP structures are of comparable magnitude 
to those found in large-scale sunspots 
\citep[$\ge1$~kG;][]{Cau00, Jes10b}, making them extremely viable 
conduits for carrying a variety of MHD waves, including fast 
and slow magneto-acoustic modes, in addition to Alfv{\'{e}}n 
waves. MBPs have demonstrated their ubiquity in both observational and 
simulated datasets, with an example of such features 
displayed in Figure~{\ref{Jess_MBP_figure}}.

The simultaneous development of theoretical, 
numerical and analytical modelling tools for the lower solar atmosphere 
has allowed the unification of 
MHD with seismological techniques 
\citep[e.g.,][]{Kim08, Ver11, Kur13, Mor14}. 
Now, observers are able to 
quantify and understand oscillatory parameters that may 
be below the resolution limit imposed by even the largest of modern 
ground-based optical telescopes (e.g., the 1.6{\,}m New Solar Telescope, 
NST, at the Big Bear Solar Observatory). This has important 
consequences, particularly when attempting to identify multiple 
modes existing within the same structure 
\citep[e.g.,][]{Mor12}, or when trying to 
diagnose small-amplitude waves that may often become 
swamped by instrumental noise and/or the point spread function 
of the telescope. In this chapter, we will review the 
recent observations of waves and oscillations manifesting in 
fine-scale magnetic structures in the solar photosphere, 
which are often interpreted as the ``building blocks'' of the 
magnetic Sun.

\section{Magneto-acoustic Waves}

The launch of the Hinode \citep{Kos07} space telescope, 
equipped with the 0.5{\,}m Solar Optical Telescope 
\citep[SOT;][]{Sue08, Tsu08}, provided researchers with 
easily accessible and high resolution ($\sim$$0.2''$ or $150${\,}km) 
observations of the lower solar atmosphere. The lack of 
atmospheric turbulence results in the images being 
`seeing free', and thus suitable for long duration studies of 
small-scale magnetic elements on the surface of the Sun. 
Notably, \citet{Car07} were at the forefront of employing 
the high resolution Ca~{\sc{ii}}~H and continuum filters on-board 
SOT to examine the propagation of acoustic 
waves down to the diffraction limit of the instrumentation. 
Through examination of all spatial locations 
(both magnetic and non-magnetic) \citet{Car07} 
speculated that the total energy flux provided, 
even at such high spatial and frequency resolutions, 
was insufficient to contribute to atmospheric heating. 
However, \citet{Wed07} stated that the methods used 
may overlook dynamic patterns created on sub-resolution 
scales, and as a result severely underestimate the actual 
mechanical flux \citep{Kal07, Kal08}. Importantly, this work 
inspired many other solar scientists to focus on highly magnetic 
photospheric elements, where the strong magnetic flux 
concentrations may promote more efficient energy propagation. 

\begin{figure*}[!t]
\begin{center}
\includegraphics[angle=0,width=13cm]{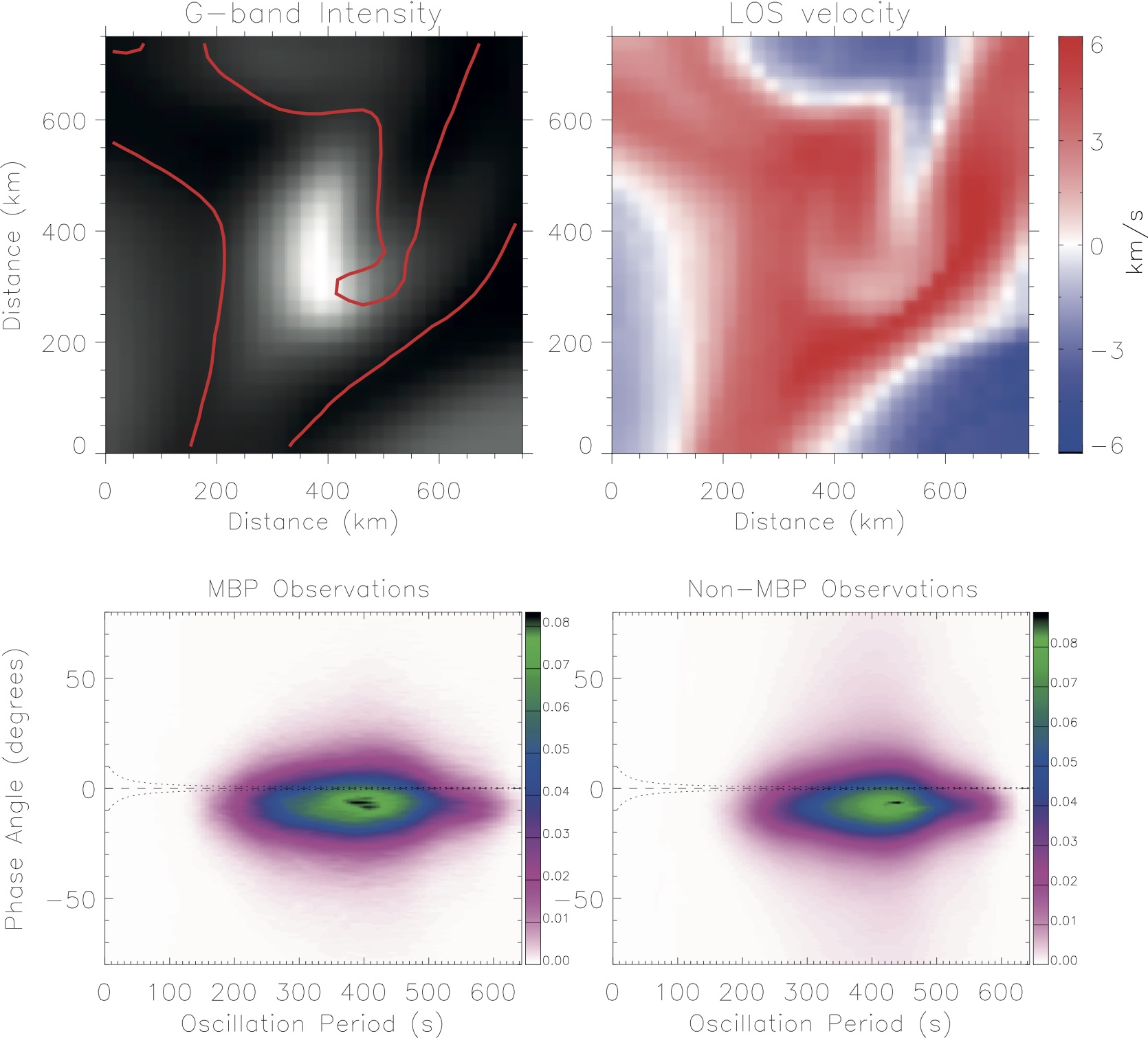}
\end{center}
\caption{Simultaneous G-band intensity (upper left) and line-of-sight velocity (upper right) 
sub-fields extracted from a synthesised 
field-of-view created using the MuRAM radiative magnetohydrodynamic 
code. Each image covers less than 1~square arcsecond, 
and is centred on a recently formed MBP structure manifesting within the 
intergranular lanes of the surrounding convective plasma. The 
red contours in the upper-left panel outline locations where the downflow 
(i.e., red-shifted) velocities exceed 3~km{\,}s$^{-1}$, as defined by the 
graduated colour spectrum displayed at the right-hand side of the Figure. 
The red-shifted plasma encompasses 
both the MBP and the surrounding intergranular lanes, indicating 
the process of convective collapse is likely to play an important 
role in the formation of MBP features. Note that the downflow velocities 
at the centre of the MBP, where wave phenomena is likely to 
manifest as a result of the increased magnetic field strengths, 
are slightly weaker than at its perimeter. 
The lower panels display the occurrence of oscillations simultaneously visible in 
ROSA G-band and 4170{\,}{\AA} continuum images, as a function of the 
oscillation period and phase angle for regions containing MBPs (lower left) and 
those without (lower right). The colour scale represents the 
number of detections as a percentage of the total events 
with an associated coherence level exceeding 85\%. 
A horizontal dashed line represents a phase angle of 0 degrees, while dotted lines 
highlight a region inside which detections become unreliable due 
to cadence restrictions (0.528~s for the observational time series). A preference for 
negative phase shifts highlights the abundance of upwardly propagating 
wave motion in non-magnetic and magnetic photospheric 
features, including the small-scale MBPs shown in the upper panels. Upper images 
courtesy of R.L. Hewitt, and based on the data 
presented by \citet{Hew14}, while the lower panels have been adapted from 
\citet{Jes12b}.}
\label{Hewitt}
\end{figure*}

Employing high spatial and temporal resolution observations from 
the ground-based DST, \citet{Jes07} 
detected more prevalent photospheric oscillations in what was 
believed to be magnetic concentrations surrounding a large sunspot. 
These oscillations demonstrated power well in-excess of the 
quiescent background, but unfortunately no direct information on the 
associated magnetic fields were available, and thus the waves were 
tentatively tied to magneto-acoustic phenomena. Further 
work by \citet{And07} found that the locations of small-scale 
oscillatory power in the photosphere correlated well with 
red-shifted velocities (i.e., downflows). This is in agreement with the 
theory that many small-scale magnetic elements in the photosphere 
are formed via the process of convective collapse \citep{Spr76}, which 
is further substantiated by the modern-day observations and 
MHD simulations of MBPs \citep[e.g.,][]{Utz13, Hew14}. However, 
it must be stated that while such oscillations have been tied to 
MBPs displaying red-shifted (i.e., downflow) velocities, this does 
not necessarily mean that the embedded waves are also downwardly 
propagating. A bulk motion may exist within the confines of the 
magnetic flux tube, hence giving rise to red-shifted Doppler 
signatures, yet the phase velocity of the wave may have a 
sufficiently large upward magnitude such that the overall group 
velocity, which crucially describes the velocity at which 
energy is propagated by the wave, is directed upwards.
For example, the observational work of \citet{And07} and 
\citet{Nar11}, corroborated by the numerical modelling presented by 
\citet{Hew14}, established downflow 
velocities on the order of a few km{\,}s$^{-1}$ within the confines 
of small-scale photospheric magnetic elements 
(see, e.g., the upper panels of Figure~{\ref{Hewitt}}). These subsonic 
plasma flows may easily become overshadowed by 
upwardly propagating magneto-acoustic phase velocities 
that are close to the photospheric sound speed 
\citep[$\sim$10~km{\,}s$^{-1}$;][]{Jes12b}. As a result, 
the group velocity, and therefore the direction of energy propagation, 
would be directed upwards. Recently, \citet{Kat11} further revealed how the processes intrinsic to 
convective collapse may also drive magneto-acoustic wave phenomena in small-scale 
magnetic elements, irrespective of the presence of sub-surface $p$-mode oscillations. 
The authors employed radiative MHD simulations and demonstrated 
how the coupling between external downdrafts in the intergranular lanes 
and the motions of the embedded plasma relies heavily on the inertial forces 
that act on the magnetic flux concentration. These forces act to `pump' 
the internal atmosphere of the magnetic flux tube in a downwards direction, 
which eventually causes the atmosphere to rebound, producing upwardly 
propagating magneto-acoustic waves along the magnetic field lines \citep{Kat11}.

Through examination of the magneto-acoustic wave dynamics associated 
with large-scale sunspots, \citet{Nag07} employed Hinode/SOT 
observations to reveal how oscillatory power is often drastically 
reduced in the presence of strong magnetic field concentrations; 
a common phenomenon now referred to as `acoustic power suppression' 
\citep[e.g.,][]{Woods81, Tho82, Tit92, Par07, Cho09, Ilon11, Cou13}. 
\citet{Law10, Law12} and \citet{Chi12} were able to corroborate these general findings through 
examination of Hinode/SOT G-band image sequences containing MBPs. However, 
interestingly \citet{Chi12} found evidence to suggest that magneto-acoustic 
power at the highest temporal frequencies (i.e., periodicities less than 100~s) 
actually demonstrated power amplification. Follow-up work, employing the 
Rapid Oscillations in the Solar Atmosphere \citep[ROSA;][]{Jes10c} 
multi-camera imaging system on the DST, revealed how magneto-acoustic 
waves with periodicities below 100~s demonstrate significant 
turbulent components within their power spectra \citep{Law11}. \citet{Law11} 
suggest that the observed magneto-acoustic waves may 
be generated by the interaction of the magnetic field lines 
with plasma downflows (i.e., characteristic of the convective collapse process) that are 
very turbulent in their nature.

While the observed periodicities of magneto-acoustic 
waves in MBPs generally span the entire $p$-mode spectrum, 
there is increasing evidence to suggest that the underlying 
magnitude of the magnetic field directly influences the 
dominant period. \citet{Kos13} employed the Triple Etalon 
SOlar Spectrometer \citep[TESOS;][]{Tri02} on the German 
Vacuum Tower Telescope (VTT) to obtain high-resolution 
spectroscopy of the photospheric Ba~{\sc{ii}} absorption line. 
They found that the dominant period of oscillations increases by 
15--20\% as the local magnetic field strength increases from 
500 to 1500~G. This has important implications since it 
suggests that the strong magnetic fluxes inherent to MBPs 
may be able to assist the propagation of lower frequency 
(i.e., below 3~mHz) magneto-acoustic waves into the 
chromosphere and beyond, especially if the magnetic 
field lines are suitably inclined to reduce the impact of the 
acoustic cut-off frequency \citep[i.e.,][]{Bel77}. As suggested by 
\citet{DeP04}, \citet{deW09} and \citet{Sta11}, to name but 
a few, the efficient leakage of lower frequency magneto-acoustic 
wave modes into the upper solar atmosphere may be able to 
drive a wide variety of high-temperature phenomena, including 
the oscillations observed in coronal fans, plumes and loops 
\citep[e.g.,][]{Def98, Ofm99, Ofm02, DeM03, DeM04}.

High resolution observations have clearly shown the 
existence of magneto-acoustic wave phenomena in small-scale 
photospheric magnetic elements. However, it is also of 
significant importance to determine whether these oscillations 
are propagating or standing waves from the viewpoint of 
supplying energy to the upper levels of the solar atmosphere. 
\citet{Jes12b} employed the ROSA system with blue 
continuum (4170{\,}{\AA}) and G-band filters to examine the 
propagation of waves between two discreet layers in the 
lower solar atmosphere. The MuRAM radiative 
magnetohydrodynamic code \citep{Vog05} was utilised to 
determine the formation heights of the two ROSA filtergrams 
through a comparison of their corresponding response 
functions. It was found that the continuum and G-band 
images were separated by $\sim$75{\,}km in height, thus allowing 
any propagation of waves between these two discreet layers 
to be investigated through phase-difference analysis. 
\citet{Jes12b} were able to detect a wealth of oscillatory 
phenomena in both continuum and G-band images. However, 
only oscillations with periodicities above $\sim$140{\,}s 
demonstrated coherent phase delays between the adjacent 
bandpasses. A $-$8$^{\circ}$ phase lag (traversing a physical 
displacement of $\sim$75~km) indicated upwardly propagating 
phase speeds on the order of 8~km{\,}s$^{-1}$. These 
velocities are similar to the expected sound speed 
(lower-right panel of Figure~{\ref{Jess_MBP_figure}}), 
and thus demonstrated the linear nature of magneto-acoustic 
wave phenomena in the lower solar atmosphere. 
Furthermore, as revealed in the lower panels of Figure~{\ref{Hewitt}},
the authors determined that 76\% of all MBP structures 
demonstrated upwardly propagating magneto-acoustic 
wave signatures, helping to explain why the outer regions 
of the solar atmosphere are so ubiquitously populated with 
MHD wave phenomena. 
Indeed, extensive work has recently 
been implemented to uncover the connectivity between 
lower atmospheric propagating waves and running oscillations 
ubiquitously observed in the solar corona 
\citep[e.g.,][]{Tom09, Jess12}.

\begin{figure*}
\begin{center}
\includegraphics[angle=0,width=16cm]{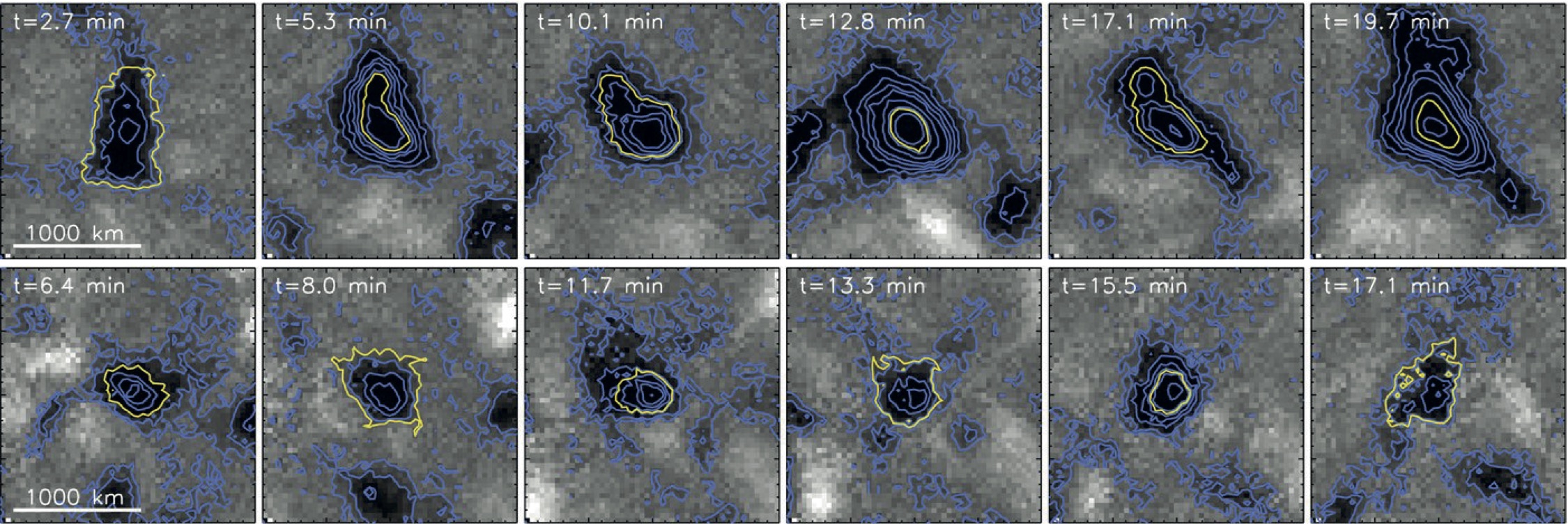}
\end{center}
\caption{Panels displaying the time evolution of circularly polarised (i.e., longitudinal) magnetic fields 
in the photosphere, as captured by the IMaX instrument on-board the Sunrise balloon flight. What may 
initially appear to be pore-sized structures are in fact sub-arcsecond magnetic concentrations 
contained within the intergranular lanes, with white and black colours representing positive and 
negative magnetic flux densities, respectively, computed using the weak-field approximation. The 
blue contours highlight various iso-magnetic flux densities, while the yellow contours represent 
time-constant magnetic fluxes equal to $-4.5\times10^{16}$~Mx (upper row) and 
$-5.0\times10^{16}$~Mx (lower row). It is clear that the time-constant iso-magnetic contours 
contract and expand with time, suggesting the presence of compressive magneto-acoustic 
waves embedded within the field lines. Images adapted from \citet{MarGon11}.}
\label{Martinez_Gonzalez_combined}
\end{figure*}

More recently, \citet{And13} employed broadband 
TiO images obtained using the NST to examine the connection between 
photospheric oscillations and the dynamic motions of 
small-scale magnetic flux concentrations. The TiO filter used 
is centred on an absorption band of molecules around 7056.8{\,}{\AA} 
(incorporating a 10{\,}{\AA} filter width), and thus
averages over all inherent absorption and continua 
contributions, causing the resulting images to be only weakly dependent 
on the properties of individual spectral lines. Ultimately, this 
means that the intensity time series will be representative of 
the true solar continuum. Furthermore, since the bandpass is 
approaching the near-infrared, the images will be less sensitive 
to atmospheric seeing variability, and thus provide better 
temporal coverage of time series obtained during mediocre 
weather conditions. The authors undertook Fourier and Hilbert 
transformations of the TiO time series related to MBPs, in order to 
extract the amplitude and phase relationships of the embedded oscillations, 
and suggested that the detected wave motion is likely to be too 
complex to be generated by a single oscillatory source. 
Nevertheless, \citet{And13} provided direct evidence for the 
presence of upwardly propagating wave trains in the immediate 
vicinity of red-shifted (i.e., downflowing) material, suggesting 
the phase velocities of the magneto-acoustic wave phenomena 
were significantly higher than 5~km{\,}s$^{-1}$ at the 
photospheric layer.

\subsection{Sausage Waves}

Even though the sausage mode is the lowest azimuthal order 
compressible mode (i.e., $m=0$), it has still proved extremely 
difficult to identify in observations. Through mathematical 
understanding and numerical modelling, these waves will demonstrate 
observational characteristics consistent with the simultaneous periodic intensity and 
area fluctuations of the magnetic flux tube. One 
of the main obstacles, at least observationally when attempting to detect 
sausage-mode oscillations, is a combination of the instrumental spatial 
resolution (i.e., to be able to detect the fractional area changes) in addition 
to the detector sensitivity (i.e., to be able to extract the small-scale intensity 
fluctuations over the intrinsic background noise). Only with the consistently 
high resolving power of modern solar facilities, coupled with the low noise 
characteristics synonymous with cooled CCD and CMOS detectors, has it 
been possible to detect sausage waves in small-scale photospheric magnetic 
elements. 

\citet{Fuj09} revealed the true power of the SOT 
on-board the Hinode spacecraft 
by examining the intensity and velocity oscillation 
characteristics in relation to the vector magnetic field 
(see, e.g., the lower panels of Figure~{\ref{Morton_combined}}). 
The authors found, through phase relationships between the various 
waveforms, that small-scale MBPs in the photosphere 
demonstrated signatures of specific magneto-acoustic 
waves, in particular the sausage and kink modes.
The observed fluctuations in the magnetic field, the line-of-sight velocity 
and the structure's intensity indicated root-mean-square amplitudes 
of $4-17$~G (0.3\%$-$1.2\%), $0.03-0.12$~km{\,}s$^{-1}$, and 0.1\%$-$1\%, 
respectively. The small amplitudes of the observed fluctuations 
emphasised the importance of high-resolution (spatial, temporal 
{\it{and}} spectral) observations when attempting to diagnose 
certain magneto-acoustic wave modes. Importantly, 
the detected oscillations maintained significant overlap with 
the global $p$-mode periodicities, further confirming the 
assumption that the majority of wave motion found in the solar 
atmosphere is driven by the underlying and 
omnipresent $p$-mode oscillations. 

Employing the spectropolarimetric capabilities of the 
Imaging Magnetograph eXperiment \citep[IMaX;][]{Mar11} 
oboard the Sunrise balloon flight, \citet{MarGon11} established 
clear evidence for periodic fluctuations in the 
magnetic field strength associated with small-scale 
($\le$$1''$) elements within the intergranular lanes. The 
isocontours related to specific magnetic fluxes were 
observed to oscillate in both area and magnitude, with 
some antiphase behaviours suggesting the presence of 
sausage-mode waves \citep{Fuj09}. However, the authors 
found that the periodicities of the oscillations were 
seldomly constant, and in fact often varied by several minutes 
within the same location (Figure~{\ref{Martinez_Gonzalez_combined}}). 
As a result, \citet{MarGon11} 
proposed that the wave driver may not be the expected 
$p$-mode oscillations, and instead may be a consequence 
of granular ``forcing''. It was proposed that since the 
average magnetic fields contained within the 
intergranular lanes may have strengths lower than 
the photospheric equipartition field 
\citep[$\sim$$300-500$~G;][]{Lin95, Kho03, MarGon08}, 
the continuously buffeting nature of granular flows may 
directly impose periodic field amplification and weakening 
through the processes associated with plasma forcing. 
Importantly, as raised by \citet{MarGon11}, is the question 
as to whether-or-not these fluctuations (regardless of 
whether they are driven by underlying $p$-modes or 
by granular forcing) are able to propagate 
upwards through the solar atmosphere, thus facilitating 
the relocation of energy to higher atmospheric heights. 
Utilising other magnetically sensitive absorption lines, 
particularly those originating in the chromosphere 
(e.g., the Ca~{\sc{ii}} infrared triplet at 8542{\,}{\AA}), will 
allow such magnetic fluctuations to be tracked through 
the solar atmosphere and shine new light on whether they 
can act as an efficient energy conduit.

Inspired by the results of 
\citet{Fuj09}, \citet{Mor13a} and \citet{Mor13b} developed 
stringent phase relationships that allow the characterisation 
of fast/slow, body/surface and standing/propagating 
sausage-mode waves based on 
the measured delays between the intensity, cross-sectional 
area and velocity components of the plasma. These relationships 
were successfully applied to the observational work of 
\citet{Mor11}, which demonstrated 
their accuracy and suitability for interpreting particular 
sausage mode properties, e.g., standing/propagating, fast/slow, 
body/surface. However, the work of \citet{Mor11} examined solar pores, which 
with diameters $>$1250~km are substantially larger than the 
$\sim$220~km sizes associated with the smallest scale magnetic 
elements in the photosphere \citep{Cro10}. Consequently, efforts 
are now being directed towards high resolution observations of 
MBPs, which will allow the phase relationships to be tested more thoroughly 
for the smallest magnetic structures currently resolvable. 

\begin{figure*}[!t]
\begin{center}
\includegraphics[angle=0,width=16cm]{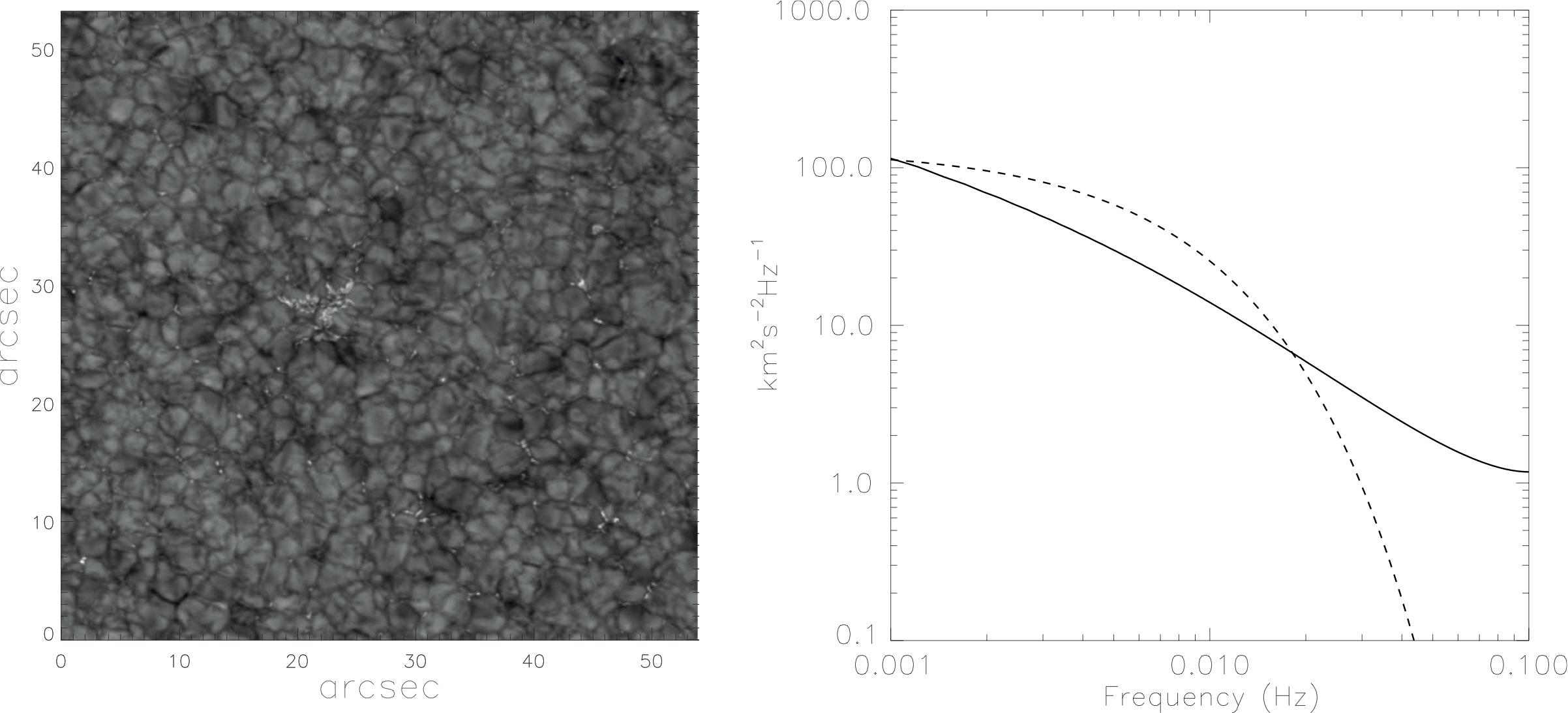}
\end{center}
\caption{A broadband H$\alpha$ image (left), acquired by the SST 
on 2006 June 18, revealing a large assortment of MBPs within 
the intergranular lanes. Following correlation tracking on each of the 
97 detected MBPs within the field-of-view, the resulting power spectrum of the 
horizontal motions is displayed as a function of frequency (solid black line) 
in the right-hand panel. The dashed line represents a standardised 
Lorentz profile using identical free parameters (e.g., the correlation time) 
to that measured in the SST observations. It is clear that for frequencies 
$>$0.02~Hz ($<$50~s) the observational horizontal motions have more 
power, highlighting the fact that dynamics on short timescales may be very 
important in the generation of energetic kink and/or Alfv{\'{e}}n wave 
phenomena. Images adapted from \citet{Chi12b}.}
\label{Chitta_combined}
\end{figure*}

\citet{Jes12a} examined the mode-coupling between compressible and 
incompressible waves found in MBPs, and the connection with their 
chromospheric spicule counterparts. The primary aim of this work 
was not to study sausage-mode oscillations, but instead their source. 
Using the Lare2D numerical code \citep{Arber2001} to model 
an MBP as a thin magnetic flux tube, the authors found 
that a $90^{\circ}$ out-of-phase behaviour of upwardly propagating magneto-acoustic 
waves at the photospheric layer directly incited the generation of 
sausage mode oscillations in the same flux tube. \citet{Jes12a} interpreted 
the numerical output as evidence for how velocity gradients embedded within 
the flux tube, as a result of the out-of-phase magneto-acoustic oscillations, 
cause the central axis of the magnetic fields to displace transversally. 
In addition, compressions and expansions in the 
waveguide are simultaneously induced, thus promoting the 
manifestation of both compressible 
sausage modes and incompressible kink waves at upper-photospheric heights.
The work of \citet{Jes12a} clearly shows how thin, magnetic structures 
omnipresent throughout the solar atmosphere can readily support 
sausage-mode wave generation and propagation, which 
is in agreement with the chromospheric work of \citet{Mor12}.

\subsection{Kink Waves}

The $m=1$ kink mode is unique in that it is the only value of 
the azimuthal wavenumber, $m$, that produces a transverse 
displacement of a magnetic flux tube. Hence, it is readily observed 
in over-dense solar structures with imagers of sufficient spatial and/or 
temporal resolution. The kink mode is highly Alfv{\'{e}}nic, since its 
main restoring force is magnetic tension, thereby making it only weakly 
compressible \citep[see, e.g.,][for a detailed discussion]{Goo09}. Kink 
waves have been most extensively studied in the corona following 
the launch of the Transition Region and Coronal Explorer 
\citep[TRACE;][]{Han99} spacecraft in 1998. Since 2007 they have 
also become the subject of much interest in the chromosphere due to 
their heightened visibility in off-limb spicules with Hinode. Since 
propagating kink waves in both chromospheric and coronal waveguides 
are now seen to be ubiquitous, it is widely believed they are being driven 
from the wealth of mechanical energy permeating the photosphere. 
As a result, kink waves have become hypothesised as a favourable 
transport mechanism for channelling energy from the photospheric 
convective motions through to the upper layers of the solar atmosphere 
\citep[e.g.,][to name but a few]{Cra05, Suz05, Ver07, Mat10}. Investigations 
into their possible photospheric signatures are now gaining momentum. 
There is some tentative observational evidence of the kink instability in 
penumbral filaments surrounding sunspots \citep[e.g.,][]{Ryu08, Bha12}, but thus 
far there has been no conclusive statistical studies undertaken of kink waves 
propagating along such structures. However, since penumbral filaments 
are predominantly highly-inclined and/or closed magnetic structures 
confined to the lower atmosphere they therefore do not represent the 
conduits required to transfer kink wave energy into the corona. A much 
more promising form of investigation is to track the photospheric 
motions that could be possible drivers which instigate kink waves. 
Since fine-scale chromospheric magnetic structures, such as 
spicules, fibrils and mottles, are rooted in intergranular lanes, their 
footpoints often reveal themselves in photospheric G-band filtergrams as 
MBPs \citep{Jes12a}. Tracking the horizontal velocity components of MBPs provides 
a useful proxy for detecting the transverse wave drivers that can excite 
kink modes, assuming that the inclinations of the associated magnetic 
flux tubes are not too far from the vertical at photospheric heights. 
The advent of high temporal and spatial resolutions from modern 
facilities has allowed observers to measure such small-scale transverse 
exertions, and as a result, the quest is now on to relate such photospheric 
horizontal velocity power spectra with that of kink oscillations observed 
higher up in the chromosphere and corona. A key scientific goal is now to 
provide an all-encompassing understanding of kink wave excitation, 
propagation and damping throughout the entire solar atmosphere.  

\begin{figure*}[!t]
\begin{center}
\includegraphics[angle=0,width=16cm]{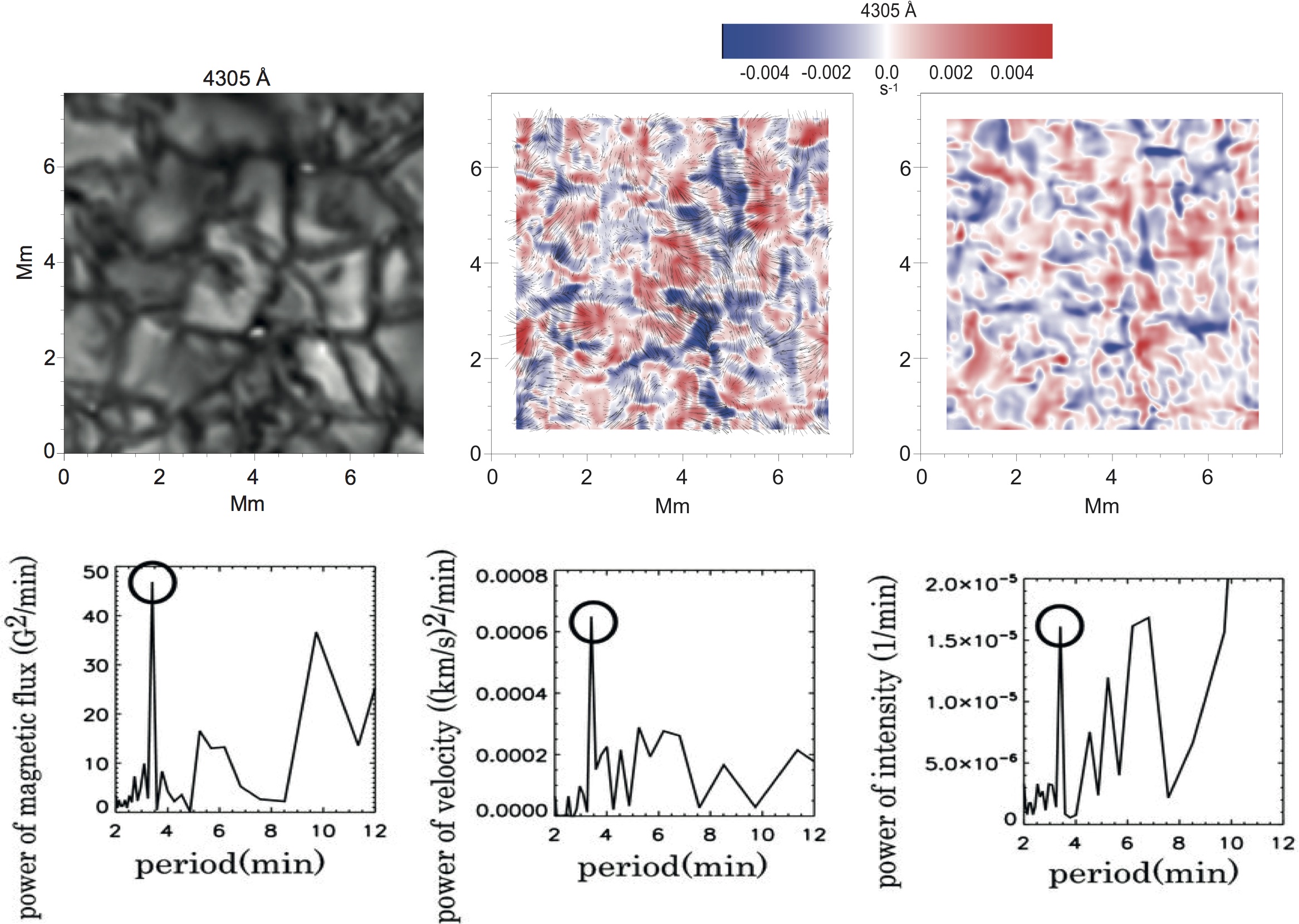}
\end{center}
\caption{A small sub-field of a ROSA G-band image (upper left), obtained 
using the DST on 2010 September 29. MBPs can be seen as 
intensity enhancements within the intergranular lanes. The upper-middle panel 
displays the simultaneous velocity vectors and divergence of photospheric 
flows. The arrows indicate the averaged velocity vectors determined 
from the local correlation tracking of G-band image sequences. The 
upper-right panel displays vorticity (s$^{-1}$, where, e.g., 0.002~s$^{-1}$ represents 
$\approx$0.11$^{\circ}$~s$^{-1}$ in the clockwise direction) calculated from 
the derived horizontal photospheric flows. It is clear that the atmosphere 
in which MBPs are embedded is replete with significant horizontal 
flows {\it{and}} torsional motion. The lower panels, from left-to-right, display 
Fourier power spectra of the line-of-sight magnetic flux, the line-of-sight velocity and 
the intensity of small-scale magnetic elements (including MBPs that are prevalent 
in the upper-left panel) observed by the 
SOT on-board Hinode. The black circles indicate narrow (common) peaks that 
represent the observational evidence of magneto-acoustic waves in the lower 
solar atmosphere, including sausage modes, which highlight 
the ubiquitous nature of wave phenomena across 
all observational datasets. 
Images have been adapted from \citet{Morton13} and \cite{Fuj09}.}
\label{Morton_combined}
\end{figure*}

In a series of recent observational papers, \citet{Sta13a, Sta13b, Sta14} 
employed data, acquired by the Hinode/SOT and the 
Sunrise/IMaX instruments, to examine the interactions 
between MBPs and their surrounding granular environment 
in an attempt to uncover how transverse waves in the 
photosphere are generated. Firstly, \citet{Sta13a} found 
an abundance of photospheric magnetic elements demonstrating 
buffeting-induced transverse oscillations with periodicities under 
100~s and velocity 
amplitudes of the order of 1$-$2~km{\,}s$^{-1}$. The authors 
interpreted their results as evidence for kink waves being 
generated by granular buffeting, which are accompanied by 
longitudinal magneto-acoustic oscillations being generated 
via non-linear interactions. The results put forward by these 
authors corroborates numerous theoretical and numerical 
studies that suggest that such motions are a natural response 
of the magnetic field to external plasma forcing 
\citep[e.g.,][to name but a few]{Rob83, Ste98, Has03, Mus03, Kho08, Fed11a, Mor14}.
Of particular note is the related observational work of \citet{Key11}, who 
found that, at any one time, approximately 6\% of photospheric MBPs display 
transversal velocities exceeding 2~km{\,}s$^{-1}$, which 
according to the mathematical analyses of 
\citet{Cho93a, Cho93b} is more than sufficient to effectively and 
efficiently drive kink-mode oscillations in photospheric magnetic 
elements. 

\citet{Chi12b} employed broadband 
H$\alpha$ images, acquired by the SST with a cadence of 5~s, to 
calculate the flow velocities of MBPs, relative to their local environment, 
through correlation tracking algorithms. The broadband nature of the 
H$\alpha$ filter (8{\,}{\AA} full-width half-maximum) resulted in the 
images being dominated by photospheric contributions, as can be seen 
in the left panel of Figure~{\ref{Chitta_combined}}. The authors were able to detect, 
track and analyse 97 individual MBP features, which allowed them to 
calculate the power spectrum of horizontal (i.e., transverse) fluctuations 
at the photospheric level. When the observational power spectrum was 
compared to a standardised Lorentzian model (see, e.g., the right panel 
of Figure~{\ref{Chitta_combined}}), it was found that the Lorentzian 
power spectrum grossly underestimated the power originating within 
oscillations $>$0.02~Hz (i.e., $<$50~s periodicity). This work suggests that 
MBP dynamics on short timescales may be very 
important in the generation of highly energetic kink and/or Alfv{\'{e}}n wave 
phenomena. Furthermore, utilising both ground- and 
space-based observatories, 
\citet{Morton13} presented observations that revealed how kink waves 
can also be excited by the vortex motions of a strong magnetic flux 
concentration in the solar photosphere (Figure~{\ref{Morton_combined}}). 
The authors detected considerable horizontal flows, in addition to 
evidence for torsional vorticities, in the high-resolution observational ROSA 
data and suggested that these may instigate considerable wave motion 
in the lower atmosphere and beyond. To test their hypothesis, \citet{Morton13} 
computed photospheric flow vectors from complementary MuRAM simulations, 
which also indicated that small vortical movements of the photospheric plasma, with 
magnitudes up to $\sim$0.3$^{\circ}$~s$^{-1}$, can help generate the 
kink waves observed in both high-resolution simulations and observations.
With MBPs covering an estimated 2.2\% of quiet 
Sun locations \citep{Sanchez10}, it does not seem inconceivable 
that the generation of kink motions in such small-scale magnetic 
fields may be responsible for the delivery of significant energy to 
higher atmospheric layers.

Taking this one step further, and 
employing the long-duration observations provided by the 
spaceborne Hinode/SOT instrument, \citet{Sta13b} examined 
the spectral characteristics (in the Fourier domain) of small-scale 
magnetic elements undergoing kink-like oscillations. Interestingly, 
the authors found that while the majority of the transverse oscillatory 
periods were in the range of 1$-$12~mHz (85$-$1000~s), there 
was no specific features unifying the wave phenomena originating within 
different magnetic structures. As a result, the authors concluded that 
the spectral characteristics represented a unique signature of each 
magnetic element itself rather than an overarching relationship that 
defines a collective of small-scale magnetic structures. 

Most recently, 
\citet{Sta14} employed a long time series of high-resolution photospheric 
magnetograms to study the effects of turbulent convection on the 
excitation of kink oscillations in small-scale magnetic elements. Importantly, 
the authors utilised empirical mode decomposition techniques, which allowed 
them to more accurately analyse non-stationary time series which may be 
dominated by the horizontal displacements of magnetic flux tubes that 
are continuously advected and dispersed by granular flows. Sub-harmonics 
of fundamental kink oscillations, with periodicities of $7.6\pm0.2$~min, were 
used to verify the hypothesis that kink waves are induced through the 
buffeting of magnetic field lines lying at the boarders of photospheric 
convective cells. An important aspect of this work is the potential for 
such kink oscillations to be excited via non-linear interactions. The 
presence of period-doubling cascades in the observational results can 
be interpreted as a signature of chaotic excitations in non-linear 
systems \citep[e.g.,][]{San09, San10}. This has important implications for 
the generation of a wide spectrum of viable propagating periodicities in 
the lower solar atmosphere, 
particularly if the driving (i.e., buffeting) frequencies are at values 
that are not necessarily multiples or fractions of the flux tube's 
natural frequency. Therefore, the work of \citet{Sta14} tentatively 
suggests that a broad spectrum of wave periodicities in small-scale 
magnetic elements can be generated regardless of the input 
motions. 

\section{Alfv{\'{e}}n Waves}

Modern MHD simulations of the lower solar atmosphere 
clearly show how torsional motions can easily be induced in 
magnetic elements in the photosphere through the processes 
of vortical motions and/or buffeting by neighbouring granules 
\citep{Mat10, Fed11b, Vig12}. 
The theoretically driven work of \citet{van11} and \citet{Asg12} 
suggested that random displacements of the photospheric 
anchor points of the magnetic field lines, with velocities on the order of 
1.5~km{\,}s$^{-1}$, would be sufficient to induce significant 
wave turbulence, thus potentially creating an efficient dissipation 
mechanism for Alfv{\'{e}}n waves. \citet{Chi12b} provided indirect 
evidence of this effect by comparing the velocity correlation 
functions of small-scale magnetic elements in the photosphere. 
The authors detected significant power associated with 
high-frequency ($>$0.02~Hz; $<$50~s) horizontal motions, and 
suggested that these cases may be especially important in the 
creation of a turbulent environment that efficiently promotes 
Alfv{\'{e}}n wave dissipation.

Aside from the numerous theoretical studies related to 
Alfv{\'{e}}n wave generation and dissipation, and as 
documented by \citet{Mat13}, observationally identifying 
pure torsional Alfv{\'{e}}n waves in the solar atmosphere has been a 
monumental struggle ever since they were postulated by 
\citet{Alf42}. In their most simplistic form (i.e., the torsional 
Alfv{\'{e}}n wave with azimuthal wave number $m=0$), their 
incompressible nature provides the inability 
to detect them through typical intensity (i.e., density) measurements, 
and the azimuthal nature of the oscillation provides no transversal 
deflection of the magnetic structure about its central axis. 
Furthermore, the narrow nature 
of magnetic flux tubes in the lower solar atmosphere causes 
difficulties when attempting to resolve the intrinsic blue- and 
red-shifts associated with velocity measurements at 
opposite edges of the structure. However, the torsional motion 
of a magnetic element carrying an Alfv{\'{e}}n wave will induce a 
degree of non-thermal line broadening when observed 
using spectroscopic techniques \citep{Zaq03}. The magnitude of 
the broadening will depend on the velocity amplitude of the 
torsional mode, which may also be compounded by bulk flows 
and/or turbulence embedded in the plasma. 

\begin{figure*}[!t]
\begin{center}
\includegraphics[angle=0,width=16cm]{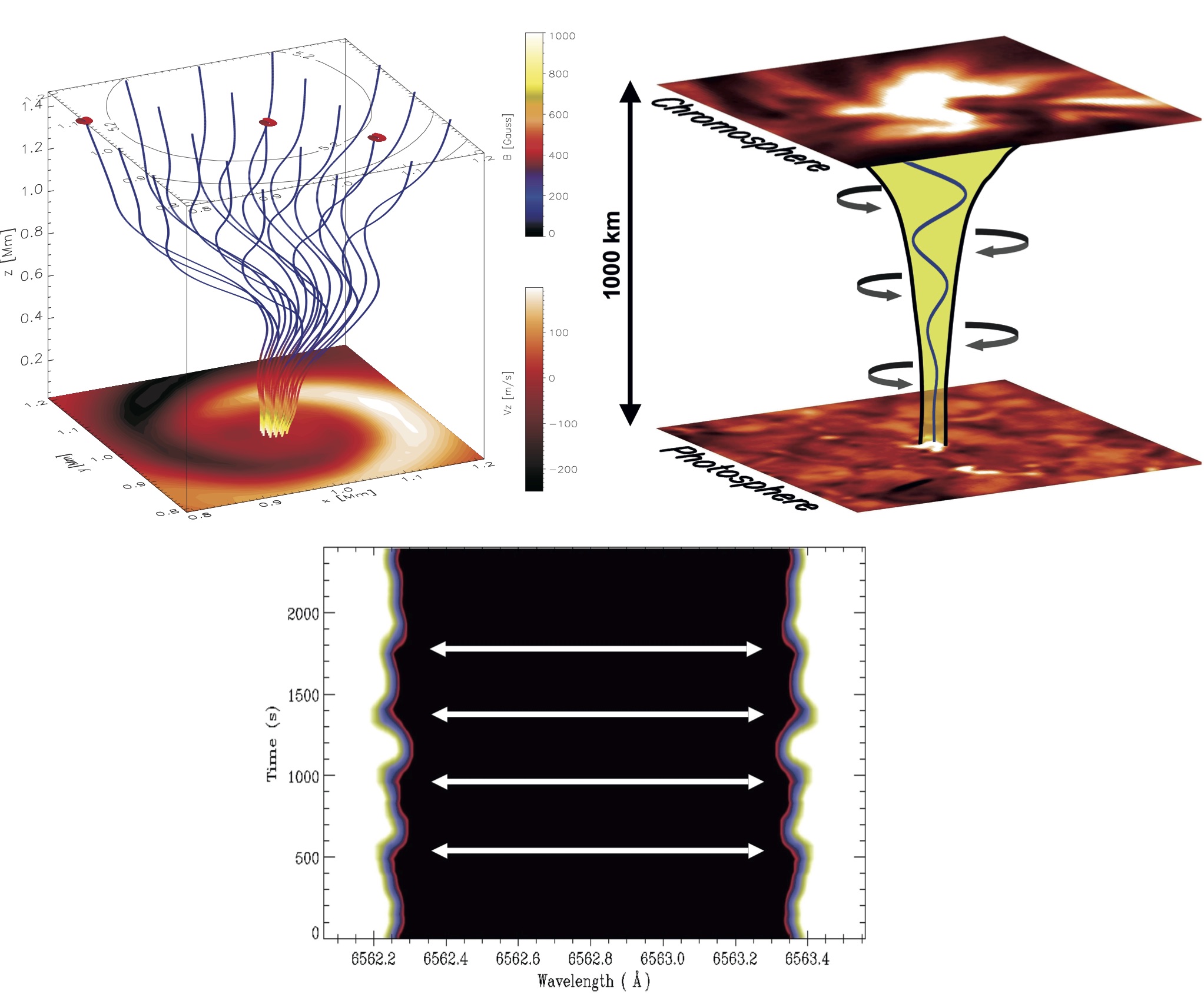}
\end{center}
\caption{The upper-left panel displays a three-dimensional snapshot of MHD 
wave propagation in an open magnetic flux tube. The simulated 
image is obtained using the nonlinear magnetohydrodynamic SAC 
code, where the thin multicolor 
curves represent the magnetic field lines that are scaled using typical 
field strengths synonymous with MBPs (i.e., $0 - 1000$~G). The upper 
and lower colour bars correspond to the magnitude of the magnetic field 
and the vertical velocity, $V_{z}$, at the level of the photospheric driver, 
resepctively. Iso-contours of the magnetic field are displayed as solid 
black lines which are labelled in the top horizontal slice taken at 
an atmospheric height of 1.4~Mm (or 1400~km). The bottom of the 
three-dimensional grid displays a horizontal cross-cut through the 
location of the torsional driver. The upper-right panel displays a typical 
expanding magnetic flux tube sandwiched between photospheric 
(broadband H$\alpha$) and chromospheric (narrowband H$\alpha$) 
intensity images obtained with the SST on 2007 August 23. 
The observational 
dataset revealed that the magnetic flux tube underwent a torsional
Alfv{\'{e}}nic perturbation, indicated by the periodic non-thermal 
spectral broadening displayed in the lower panel. In the wavelength-versus-time 
plot the H$\alpha$ absorption profile line width, calculated using spectral imaging 
techniques, is observed to 
oscillate with a periodicity $\sim$420~s, with consecutive peaks indicated by the white 
arrows. As can be seen in both upper 
panels, the Alfv{\'{e}}nic displacements are torsional motions that
remain perpendicular to both the direction of propagation and 
the magnetic fields outlining constant magnetic surfaces. 
Images adapted from \citet{Jes09} and \citet{Fed11b}.}
\label{Fedun_Jess_combined}
\end{figure*}

Employing the high-resolution spectral imaging capabilities 
of the Solar Optical Universal Polarimeter 
\citep[SOUP;][]{Tit86} on the SST, 
\citet{Jes09} were able to identify periodic non-thermal line 
broadening associated with a torsional Alfv{\'{e}}n wave 
embodied in the magnetic field lines anchored into 
a conglomeration of photospheric MBPs 
(Figure~{\ref{Fedun_Jess_combined}}). Importantly, their 
interpretation was further strengthened by the fact that 
the magnetic element did not display any periodic 
fluctuations in intensity or longitudinal/transverse 
velocity, which helped verify the absence of other 
magneto-acoustic modes. In follow-up work, 
\citet{Mat13} revealed the independent and opposite 
Doppler shifts associated with the opposing edges of the magnetic 
element, thus reinforcing the interpretation that the 
observational signatures represented a torsional Alfv{\'{e}}n 
wave. In this work, it was found that the amplitude of the 
non-thermal broadening was $\sim$0.05~{\AA}, equating to 
a velocity amplitude of $\sim$2.5~km{\,}s$^{-1}$. When 
combined with a local Alfv{\'{e}}n speed on the order of 
22~km{\,}s$^{-1}$ and a plasma density of approximately 
$1\times10^{-9}$~g{\,}cm$^{-3}$, the resulting wave 
energy was estimated as $\sim$150{\,}000~W{\,}m$^{-2}$. 
While this is a vast quantity of available energy, far in excess 
of the threshold required to sustain localised chromospheric and 
coronal heating, the true importance can only be ascertained 
once the filling factor of such waves is accurately known 
(see, e.g., Chapter~24). 
However, as documented by 
\citet{Goo11} and \citet{Mat13}, Alfv{\'{e}}n waves are 
naturally difficult to dissipate unless they are able to 
find an alternative mechanism to promote efficient 
energy dissipation. Such mechanisms include 
phase mixing and resonant absorption 
\citep[e.g.,][]{Goo01}, whereby non-uniformities in the 
magnetic field configurations results in the coupling 
between neighbouring magnetic iso-surfaces, thus promoting 
a significantly steep gradient to allow efficient energy 
dissipation. Other possibilities are the mode conversion of 
Alfv{\'{e}}n waves into magneto-acoustic modes that can 
propagate obliquely to the magnetic field lines, thus promoting 
efficient energy loss \citep[e.g.,][]{Par91, Nak97}, or turbulent 
mixing as a result of high frequency fluctuations 
\citep[e.g.,][]{van11, Asg12}.

Utilising the same dataset as presented by \citet{Jes09}, 
\citet{Fed11b} examined the spatial structuring of the 
observed torsional Alfv{\'{e}}n frequencies and related these 
to the outputs of nonlinear three-dimensional 
magnetohydrodynamic numerical simulations from the 
Sheffield Advanced Code \citep[SAC;][]{She08}. The authors 
implemented a vortex driver at the base of the simulated domain 
(see the left-hand image of Figure~{\ref{Fedun_Jess_combined}}) 
and revealed how magnetic flux tubes can act as a spatial 
frequency filter for torsional Alfv{\'{e}}n waves. Importantly, 
the authors found that his form of frequency filtering is strongly 
dependent on the structure and geometry of the magnetic field itself. 
This implies that the observed spatial wave power and oscillatory 
frequencies can be a function of the underlying MBP, possibly 
allowing magnetic fields to be mapped as a function of atmospheric 
height solely employing such seismology techniques \citep{Fed11b}. 

Even with higher sensitivity equipment now becoming 
commonplace on a majority of ground-based telescopes 
\citep[e.g., the IBIS and CRISP imaging spectropolarimeters;][]{Cav06, Sch08}, 
there is still a significant lack of subsequent Alfv{\'{e}}n 
wave detections in the lower solar atmosphere. 
This is in stark contrast to modern numerical simulations 
\citep[e.g., the MuRAM code;][]{Vog05} which indicate 
widespread torsional motions in synthesised photospheric 
filtergrams \citep{She13}. Instead, many have 
turned their attention to the chromosphere where the 
interaction, mode conversion and dissipation of 
Alfv{\'{e}}n waves may have more identifiable signatures, 
particularly in regions where the plasma $\beta$ changes 
abruptly, or self-induced turbulence results in rapid 
localised dissipation \citep{van11, Asg12}.

\section{Conclusions \& Future Directions}
\begin{figure*}[!t]
\begin{center}
\includegraphics[angle=0,width=16.5cm]{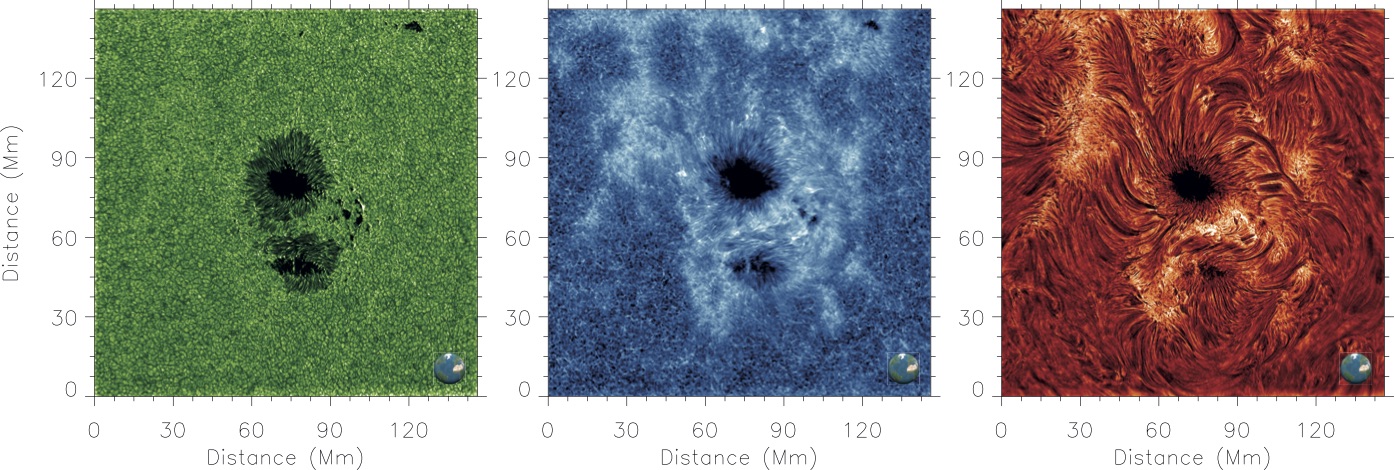}
\end{center}
\caption{Simultaneous images of a solar active region, captured 
using both existing and next-generation imaging detectors on the 
DST during 2014 August 24. The left and middle images are 
G-band (photosphere) and Ca~{\sc{ii}}~K 
(upper-photosphere/lower-chromosphere) snapshots, respectively, acquired 
using the electron-multiplying CCD cameras of the ROSA 
multi-camera system. A true-size representation of the Earth is 
depicted in the lower-right section of each image to provide a sense 
of scale. 
The image on the right reveals the solar 
chromosphere, captured through a 
$0.25${\,}{\AA} H$\alpha$-core filter, and employs a 4.2~MP 
Zyla CMOS sensor from Andor Technology, which allows image 
sequences to be obtained with 15~ms exposure times and 
frame rates exceeding 60~s$^{-1}$. Furthermore, the large pixel 
formats of modern CCD and CMOS detectors allow fields-of-view 
in excess of $200'' \times 200''$ 
to be sampled at the diffraction limit, and when combined with 
numerous detectors each sampling discreet wavelengths, provides 
a seamless view through the tenuous atmospheric layers.
Combining such high-sensitivity detectors with current 
and future telescope facilities will open the door for 
greater scientific understanding through drastically improved 
number statistics and larger fields-of-view. Images courtesy of 
D.B. Jess.}
\label{Jess_24Aug2014}
\end{figure*}

In the near future it is expected that 
high sensitivity 2D spectropolarimeters (e.g., CRISP and IBIS) will be 
employed simultaneously alongside high-cadence imagers 
(e.g., Hinode/SOT and ROSA) to obtain multiwavelength 
time series at the highest spatial, temporal and spectral 
resolutions currently achievable (see, e.g., Figure~{\ref{Jess_24Aug2014}}). 
High precision measurements 
will allow the characterisation of MHD waves themselves manifesting in 
small-scale magnetic elements that are at the 
limits of current telescope resolving power. 
Important oscillatory parameters, such as the propagation speeds, 
amplitudes and phase relationships will allow MHD wave 
phenomena to be documented with unprecedented accuracy, including the 
establishment of evidence to verify the presence of standing/propagating, 
fast/slow, trapped/leaky and surface/body oscillatory modes. 
Furthermore, the multiwavelength nature of the data will also enable 
the detected MHD waves to be tracked through the solar atmosphere 
as they journey from the photosphere, through the tenuous chromosphere, 
and into the super-heated corona. Importantly, coverage of 
the waves as they propagate through the different atmospheric regions 
where the plasma $\beta$ changes from magnetically dominated 
to plasma pressurised regimes will provide valuable insight on 
aspects of mode coupling and wave dissipation. Many previous 
examples have indicated that regions where $\beta=1$, 
often in locations sandwiched between the photosphere and 
chromosphere, provides 
opportune atmospheric conditions to promote efficient oscillatory 
mode conversion, thus allowing naturally difficult-to-dissipate 
waves (e.g., Alfv{\'{e}}n waves) to convert into more 
readily dissipated compressible modes 
\citep[e.g.,][to name but a few]{Ulm91, Kal97, Has03, McA03, Blo06, Goo06, Jes12a}.

In addition to maximising the scientific return of the current 
fleet of ground- and space-based solar telescopes, 
the solar community eagerly awaits the arrival of the first 
next-generation high-resolution facilities, including the 
$2${\,}m National Large Solar Telescope \citep[NLST;][]{Has10} in Ladakh, India, 
and the 
$4${\,}m Daniel K. Inouye Solar Telescope 
\citep[DKIST, formerly the Advanced Technology Solar Telescope, ATST;][]{Kei03, Rim10} 
atop the Haleakal{\={a}} volcano on the Pacific island of Maui. These pioneering facilities 
are due to receive first light towards the latter stages of this 
decade, and by utilising dramatically 
increased aperture sizes (4{\,}m in the case of DKIST), 
photospheric structures down to $\sim$$20$~km in size will 
be able to be detected, tracked and studied in unprecedented detail. 
In the longer term is the launch of the Solar--C payload (expected launch 
date around 2019) with a UV/visible/IR 
telescope approaching 1.4{\,}m in diameter \citep{Sue14}. The 
seeing free data from a high-resolution space-based facility 
will enable near-continuous observations of magnetic elements to be obtained 
without the need for extensive post-processing algorithms. Even more 
long-term is the 
4{\,}m European Solar Telescope \citep[EST;][]{Col10} and the 
5--8{\,}m Chinese Giant Solar Telescope \citep[CGST;][]{Liu14}, which 
are expected to begin construction following 2020, and allow structures 
as small as $\sim$$10$~km in size to be observed for the first time. 

The desire to improve not only the spatial resolution of solar telescopes, 
but also the temporal and spectral resolutions through the development 
of new imaging and spectropolarimetric instruments, will enable key 
outstanding questions related to lower atmospheric understanding to be 
firmly addressed. Such questions include:
\begin{itemize}
\item {\it{Do all small-scale magnetic elements carry energy in the form of 
MHD waves?}} There is an abundance of evidence that demonstrates 
the suitability of narrow magnetic strands as efficient wave/energy 
conduits. For example, \citet{Jes12b} found that $\sim$73\% of MBP 
structures demonstrated magneto-acoustic wave signatures. However, 
does this mean that $\sim$27\% of these features contain no 
oscillatory phenomena? Or do waves exist, but under different guises 
(e.g., the more difficult to detect Alfv{\'{e}}n modes)?
\item {\it{What is the dominant mode of oscillation present in small-scale 
photospheric magnetic features?}} The majority of research to date has 
focussed on the magneto-acoustic signatures detectable through 
intensity (i.e., density) fluctuations. However, with improved observations 
and detection algorithms \citep[e.g.,][]{Sta13a, Sta13b}, it has become 
clear that other wave modes may be present in small-scale magnetic 
elements alone or alongside their compressible counterparts. Are these 
additional modes (e.g., kink and Alfv{\'{e}}n waves) superimposed on top 
of the seemingly ubiquitous compressive oscillations? Or are they 
the dominant mode of oscillation in certain magnetic features? And if so, 
which features?
\item {\it{How do the $\beta=1$ layers contribute to both the waveforms 
visible in the outer solar atmosphere and the rate of localised heating 
through wave dissipation?}} It is well accepted that regions of the Sun's 
atmosphere where $\beta=1$ (i.e., the magnetic pressure is equal to the gas pressure) 
are optimal for oscillatory mode conversion \citep[e.g.,][]{Ulm91, Kal97, Has03}. 
\citet{McA03} and \citet{Blo06} have demonstrated the coupling between 
kink waves and longitudinal oscillations, while \citet{Jes12a} have 
documented the conversion between longitudinal waves and kink/sausage 
modes. Regions where $\beta=1$ are also predicted to have important 
consequences for Alfv{\'{e}}n wave conversion through processes 
such as resonant absorption \citep[e.g.,][]{Goo06}. Can such atmospheric 
layers convert observed (and possibly unobserved!) waves into 
other modes that can readily dissipate their energy into the surrounding 
plasma? Do the preferred conversion mechanisms substantiate 
the abundance of oscillatory motion observed in the outer regions of the 
solar atmosphere?
\item {\it{Can we find evidence for the existance of magneto-acoustic 
wave modes with azimuthal wave numbers exceeding 1?}} The observational 
work to date suggests an abundance of sausage ($m=0$) and kink ($m=1$) 
mode oscillations in the lower solar atmosphere. However, theoretical 
predictions also suggest that higher order wave numbers 
(i.e., fluting modes with $m\ge2$) 
should also be prevalent in magnetic flux tubes 
\citep[for a recent review see][and the references therein]{Goo11}. 
Why do we currently have no evidence for such oscillations? Is it a 
result of relatively poor telescope resolution and/or instrument sensitivity? Or 
are the amplitudes of these oscillatory modes so small that they become 
impossible to disentangle from other superimposed waves?
\item {\it{What role do downwardly propagating MHD waves play in the 
structuring, evolution, dynamics and energy balance of magnetic features in the 
solar photosphere?}} Attempting to find the elusive solution to the 
coronal heating problem has resulted in an overwhelming interest in 
upwardly propagating MHD waves. However, observations 
continue to identify significant oscillations propagating downwards through 
the solar atmosphere \citep[e.g.,][]{Gup13}. \citet{Jes12b} estimated that 
approximately a quarter of all detected magneto-acoustic waves in 
MBPs were downwardly propagating. Are these propagating oscillations 
generated at higher atmospheric heights? Are they the counterpart of 
upwardly propagating oscillations generated at the opposite foot point 
of a closed loop? Or are they the result of 
reflection as upwardly propagating waves encounter the severe density 
discontinuities intrinsic to chromospheric and transition region layers? 
Furthermore, how do these downward motions contribute to the mass/energy 
flow associated with magnetic elements in general?
\end{itemize}

The above questions are not intended to highlight all of the outstanding problems 
related to MHD wave phenomena in the lower solar atmosphere, nor are they 
listed in order of importance. Instead, they 
are simply listed as those which we feel have an overarching central importance 
when attempting to address the long-standing issues of wave generation, propagation, 
energy transfer and dissipation throughout the photosphere and beyond. 
Furthermore, it seems unlikely that the above questions can be unequivocally 
answered relying solely on observational approaches. A combination of 
theoretical, analytical and numerical modelling techniques will be required to 
help extract and interpret the wealth of MHD wave modes that 
exist in the lower solar atmosphere. Thus, the answers to the key science questions
outlined above will only arise through the novel use of high-resolution 
(spatial, temporal {\it{and}} spectral) 
photospheric datasets alongside the rapid development of our theoretical 
understanding.


%
%
%
%
%
%
%

\begin{acknowledgments}
D.B.J. wishes to thank the UK Science and Technology 
Facilities Council (STFC) for the award of an Ernest 
Rutherford Fellowship alongside a dedicated Research Grant. 
G.V. acknowledges the support of the Leverhulme Trust (UK).
\end{acknowledgments}

\end{article}
%
%
%
%
%
%
%
%


\end{document}